\documentclass[12pt,preprint]{aastex}

\begin{document}

\title{On the Effect of Explosive Thermonuclear Burning on the 
Accreted Envelopes of White Dwarfs in Cataclysmic Variables}

\author{Edward M. Sion             
} 
\affil{Astronomy \& Astrophysics, Villanova University, \\ 
Villanova, PA 19085, USA \\ 
\email{edward.sion@villanova.edu} 
}

\begin{abstract} 
The detection of heavy elements at suprasolar abundances in the 
atmospheres of some accreting white dwarfs in cataclysmic variables,
coupled with the high temperatures needed to produce these elements
requires explosive thermonuclear burning. The central temperatures of any
formerly more massive secondary stars in CVs undergoing hydrostatic
CNO burning are far too low to produce these elements. Evidence 
is presented that at least some cataclysmic variables contain donor 
secondaries that have been contaminated by repeated novae ejecta 
and are transferring this material back to the white dwarf. This 
scenario does not exclude the channel in which formerly more massive 
donor stars underwent CNO processing in systems that underwent 
thermal timescale mass transfer. Implications for the progenitors of CVs are discussed.
\end{abstract} 

\section{Introduction}
 
Approximately 20\% of the cataclysmic variables (CVs) reveal large 
abundance ratios of nitrogen to carbon, most commonly from the ratio 
of intensities of N\,{\sc v} (1238, 1242) to C\,{\sc iv} (1548, 1550) resonance 
doublet emission lines in the far ultraviolet \citep{gan03}.           
The N\,{\sc v} emission is typically very strong and the C\,{\sc iv} emission is 
very weak or absent. This enhancement of nitrogen and depletion 
of carbon is the hallmark of CNO processing during Hydrogen burning via 
the CNO bi-cycle. The emission lines almost certainly arise from the 
accretion disk (or boundary layer) that forms when the Roche-lobe 
filling donor secondary transfers gas to the white dwarf primary 
star. The N/C abundance anomaly has also been seen in magnetic CVs
 such as AE Aqr \citep{mou03}, BY Cam \citep{mou03}, V1309 Ori 
\citep{szk96} and MN Hya \citep{sch01},  

A less common manifestation of the N/C abundance anomaly is seen 
in a few cases where detected N and C absorption lines form in the 
exposed white dwarf photosphere itself. The two best examples are 
the white dwarfs in VW Hyi \citep{sio95,sio97,sio01} and U Gem 
\citep{sio98}. Does the N/C abundance 
anomaly arise from a formerly more massive secondary star (capable 
of CNO burning) having been peeled away by mass transfer down to 
its CNO-processed core, due to mass transfer? Or does the N/C 
anomaly originate in the white dwarf itself due to unstable CNO 
burning associated with nova explosions or possibly by dredgeup 
and mixing in consequence of a dwarf nova outburst? 
  
The question of which star, the white dwarf or the donor main sequence 
star, is responsible for the N/C anomaly is critical. If the N/C 
anomaly originated in the donor, then the donor had to be more massive than
its present value to sustain CNO burning. This could be the case 
if the CVs with the N/C anomaly are the descendants of the supersoft 
X-ray binaries, when the formerly more massive donor in the system 
underwent thermal timescale mass transfer at a high rate, thus 
driving steady thermonuclear burning on the white dwarf at the 
accretion supply rate. In this Letter, I present evidence that, 
for at least a fraction of the CVs that reveal the N/C anomaly, 
the origin of the CNO processed abundances lies with the white 
dwarf which subsequently contaminated
the secondary donor star with the ejecta of many past repeated nova explosions. 

\section{Chemical Abundances}

For the vast majority of the CVs that exhibit the N/C anomaly,
the highly non-solar abundances of N and C are inferred qualitatively 
from emission line intensity ratios and no other atomic species
are used. There have been no reported abundances of N and C determined 
ab initio from fitting emission line profiles, to date. However, 
a handful of dwarf novae, all but one below the CV period gap with 
orbital periods below 2 hours, were observed during dwarf nova 
quiescence when the accretion rate is very low and the white dwarf 
completely dominates the far ultraviolet band. These systems are 
BW Scl, SW UMa, BC UMa and VW Hyi while U Gem is the only system 
above the period gap. The chemical abundances of their accreted 
metals have been derived largely from high quality HST and FUSE 
spectra by fitting the observed metal lines with rotationally 
broadened theoretical line profiles, using TLUSTY and SYNSPEC 
\citep{hub88,hub95}. In three SU UMa-type CVs, 
BW Scl, SW UMa and BC UMa, the detected photospheric features 
reveal aluminum abundances of 
3.0$\pm$0.8, 1.7$\pm$0.5, and 2.0$\pm$0.5, respectively 
\citep{gan05}.

The dwarf nova VW Hydri's white dwarf is also detected during dwarf 
nova quiescence and is modestly hotter than the white dwarfs
in the three systems above, and has a slightly longer orbital period.
A plethora of photospheric absorption features due to metals have 
been detected in the VW Hyi white dwarf. The derived abundances 
(relative to solar values) from profile fitting are 
Al 3, Si 0.3, C 0.3, O 3, N 3.0, Al 2.0, P 20, and Mn 50 
\citep{sio95}. Large suprasolar phosphorus abundances result 
from thermonuclear runaways on ONeMg white dwarfs
\citep{sta06,jos06}.

The exposed white dwarf in the dwarf nova U Gem also exhibits a 
rich array of absorption features due to metals in both IUE and 
FUSE spectra \citep{sio95}. From HST and FUSE spectra, the U 
Gem white dwarf has the largest photospheric N/C ratio of any CV 
white dwarf \citep{sio97,lon06}.    

\section{Hot CNO Burning and the Abundances of Elements in the Mass Range A $>$ 20}

It is well known that Hydrostatic hydrogen burning via the CNO bi-cycle 
powers upper main sequence stars with masses $M >  1.3M_{\odot}$  
and the reactions are highly temperature-sensitive. It is equally 
clear that hot CNO burning depletes the abundance of carbon while 
increasing the abundance of nitrogen. Hence, the CNO burning of 
a previously more massive CV secondary star could give the large 
N/C abundance ratios that one observes in $\sim$20\% of the CV 
population. However, even in main sequence stars as massive as 
5 to 10$M_{\odot}$, the core temperatures do not exceed 
$\sim 5\times 10^{7} $K.  

The suprasolar chemical abundances of nuclides with A$>$20 that 
have been detected in the photospheres of the accreting white dwarfs 
in several dwarf novae (e.g. Al, P, Mn) present a serious problem for 
any previously more massive CV secondary. First of all, there is no 
leakage of A$>$20 elements linking the CNO reactions to A$>$20 species.
Hence, any nuclide with A$>$20, during hydrogen burning, is not 
matter that is converted from CNO material. Thus hydrostatic burning 
beyond the CNO mass range must begin from pre-existing A$>$20 nuclides. 
For example, the suprasolar abundance of ${}^{27}$Al must be formed 
in proton capture reactions starting with seed ${}^{24}$Mg 
\citep{ili07} and so on.
At temperatures below $5 \times 10^{7} $K, it is not possible for ${}^{27}$Al
to be produced from pre-existing ${}^{24}$Mg. The Coulomb barrier is simply far 
too high for such proton capture reactions.
Indeed, to build up the ${}^{27}$Al by depleting ${}^{24}$Mg by proton capture 
reactions, the temperature of a main sequence star would have to be
$\sim 7.5 \times 10{^7} $K, which corresponds to a 25$M_{\odot}$ star 
to explain the observed Mg-Al anti-correlation observed in 
globular cluster stars \citep{pra07}. 

\section{Summary of Conclusions}

The production of A$ > $20 nuclides during hydrogen burning
requires temperatures between 100 million and 400 million degrees K
to affect the needed transmission through the Coulomb barriers of 
these heavier nuclei. The only thermonuclear environment that reaches 
these temperatures is during explosive hydrogen burning, e.g. a 
classical nova thermonuclear runaway. Hence, the detection of 
such nuclei as P and Al at suprasolar abundance in the surface 
layers of accreting white dwarfs in dwarf novae implies their 
origin within the white dwarf. In the absence of mixing of the 
accreted material with the nucleosynthetic products in the 
transition region at the base of the accreted envelope where 
thermonuclear burning occurs, the matter flowing over from the 
secondary star to the white dwarf would cover its surface layers. Hence,
for some fraction of the CVs exhibiting the N/C anomaly, the donor 
secondary star's large N/C abundance ratio arose from repeated 
nova-contamination. This conclusion is supported by the following considerations.

(1) During the hydrostatic hydrogen burning of a 2 to 3 $M_{\odot}$ 
main sequence star, the central temperature, while on the main 
sequence, reaches approximately $2.2 \times 10{^7}$ K generating 
nuclear energy via the CNO reactions which convert hydrogen into 
helium. Thus, since the most likely mass range for a formerly more 
massive CV secondary is 2 to 3 $M_{\odot}$, the core temperatures 
of such stars are far too small to
allow the formation of nuclei in the mass range A$  \ge $20. Thus, 
the detection of odd-numbered nuclides like Al, P and Mn implies that the
white dwarf, not the secondary, is responsible for these nuclides.
Since these heavy nuclei were accreted, the implication is that 
the secondary star, at least in the dwarf novae that are cited above, 
was contaminated by repeated nova ejecta and is transferring this 
nova-polluted material back to the white dwarf via Roche lobe overflow. 

(2) IR spectroscopy of VW Hyi's secondary star appears to have solar 
abundance \citep{ham11} but the accreting white dwarf in 
the system, detected during quiescence, has a large N/C ratio in 
its surface layers \citep{sio95,sio97,sio01}. If the IR 
spectroscopic result of \citet{ham11} is confirmed, then 
this result suggests that the large N/C ratio originated in the 
white dwarf, the same origin as the nuclides in the mass range A$>$20.  
  
(3) The FUSE spectra of U Gem's white dwarf photosphere reveals
absorption features of P\,{\sc v} during quiescence, weeks after a dwarf 
nova outburst would have dredged it up. Phosphorus is a nuclide 
that should be produced in high abundance at the very high temperatures 
of classical nova explosions and is seen in the surface layers 
of the white dwarfs in both VW Hyi and U Gem. In the atmosphere of 
U Gem's white dwarf, with a quiescent surface temperature of $\sim 
30,000$K, the diffusion timescale for a phosphorus ion is shorter 
than 3 days \citep{paq86}. Hence, the photospheric P\,{\sc v}  
had to be accreted from the secondary. Thus, this implies that the 
donor star was contaminated by material that underwent explosive 
CNO burning of elements beyond A$>$20. The secondary star in U Gem 
has a large N/C ratio \citep{ham11}. While it cannot be 
excluded that the large N/C ratio originated in its formerly more 
massive secondary star, this would mean that the detected nuclide 
${}^{31}$P originated in the white dwarf but that the large N/C 
ratio originated separately in the secondary star. An appeal to 
Occam's razor would suggest this to be less likely since two 
separate thermonuclear processes would be involved instead of only one.

(4) To date, no pre-CV white dwarf or pre-CV secondary star has yet 
been found to have surface abundances indicative of prior CNO processing. 
Yet it is the pre-CVs which are thought to be the immediate progenitors 
of the cataclysmic variables. 

Finally, while this work supports the idea that nova-contamination 
of the secondaries has taken place in the dwarf novae cited in this 
Letter, this channel for the production of A $>$ 20 stable nuclides 
and a large N/C ratio may be represent only a small subset of the 
CVs that show large N/C abundance ratios. The dominant channel 
may still be the thermal timescale mass transfer systems accreting 
at high rates from a formerly more massive secondary. Nevertheless, 
further investigations of how much material lands on the donor star 
in consequence of repeated nova explosions are clearly warranted.   

{\bf Acknowledgements}
I am deeply indebted to Jim MacDonald for referring my queries on 
nuclear reaction networks to the superb book by Christian Iliadis, 
and to Christian Iliadis for illuminating discussions. 
It is a pleasure to thank Warren Sparks for useful discussions about
accretion in CVs over the years. 
This work 
is supported by NASA Grant NNX13AF12G to Villanova University.

\end{document}